\documentclass[showpacs, pra,onecolumn,preprintnumbers ,amsmath, amssymb, superscriptaddress, aps]{revtex4-2}
\usepackage{xcolor}
\usepackage{color}
\usepackage{pifont}
\usepackage{bbold}
\usepackage{float}
\usepackage{tikz}
\usepackage{makecell}
\usepackage{graphicx} 
\usepackage[caption=false,justification=justified]{subfig}
\graphicspath{{Figures/}}
\usepackage{subfloat}
\usepackage{dcolumn}  
\usepackage{bm}       
\usepackage{multirow} 
\usepackage[colorlinks]{hyperref}
\usepackage{placeins}

\newcommand{\beq}{\begin{equation}}
    \newcommand{\eeq}{\end{equation}}
\newcommand{\bqr}{\begin{eqnarray}}
    \newcommand{\eqr}{\end{eqnarray}}
\newcommand{\bb}{\color{blue}}

\begin{document}

    \title{Transmission in graphene through tilted barrier in laser field }

    \author{Rachid El Aitouni}
    \affiliation{Laboratory of Theoretical Physics, Faculty of Sciences, Choua\"ib Doukkali University, PO Box 20, 24000 El Jadida, Morocco}

    \author{Miloud Mekkaoui}
    \affiliation{Laboratory of Theoretical Physics, Faculty of Sciences, Choua\"ib Doukkali University, PO Box 20, 24000 El Jadida, Morocco}

    \author{Ahmed Jellal}
    \email{a.jellal@ucd.ac.ma}
    \affiliation{Laboratory of Theoretical Physics, Faculty of Sciences, Choua\"ib Doukkali University, PO Box 20, 24000 El Jadida, Morocco}
    \affiliation{Canadian Quantum  Research Center,
        204-3002 32 Ave Vernon,  BC V1T 2L7,  Canada}

    \begin{abstract}

We study the transmission of Dirac fermions in graphene through a tilted barrier potential in the presence of a laser field of frequency $\omega$. By using Floquet theory, we solve the Dirac equation and then obtain the energy spectrum. The boundary conditions together with the transfer matrix method allow us to determine the transmission probabilities corresponding to all energy bands $E+l\hbar\omega$ $(l=0,\pm1, \cdots)$. By limiting to the central band $l=0$ and the two first side bands $l=\pm 1$, we show that the transmissions are strongly affected by the laser field and barrier. Indeed, it is found that the Klein effect is still present, a variety of oscillations are inside the barrier, and there is essentially no transmission across all bands.

    \end{abstract}

    \maketitle

\section{Introduction}

Graphene is one of the few two-dimensional materials that is made up of carbon atoms that form a hexagonal structure in the shape of a honeycomb \cite{Novoselov2004}. The thickness of which does not exceed the thickness of an atom. It has incredible mechanical, electronic, optical, thermal, and chemical properties \cite{Novoselov2005}, as well as extremely high mobility. The speed of its electrons is 300 times less than the speed of light \cite{Castro2009}. Because of its speed, the researchers consider its electrons to be massless Dirac fermions. It is one of the most powerful conductors of electricity and heat, which has started to be used in technology industries \cite{Bhattacharjee2006,Bunch2005,Berger2004}. It replaces ordinary semiconductors such as silicon to manufacture electronic components, solar panels, and touch displays.

The Hamiltonian describing the graphene properties is a relativistic Dirac Hamiltonian (with the speed of light replaced by $v_F=c/300$), resulting within the framework of the tight-binding model \cite{Castro2009}. The eigenvalue equation resolution reveals that the Brillouin zone is delimited by six Dirac points, which are represented by two non-equivalent points $K $and $K'$, each corresponding to two atoms of the pattern in the direct lattice. In addition, the energy is a linear dispersion relation in the vicinity of the Dirac points \cite{Zhang2005}, which is similar to a cone, i.e., the valence and the conduction band are tangent at the Dirac points. As a result, the electrons inside graphene can easily jump from the valence band to the conduction band. This is the problem that delays its use. The zero gap between these two bands prompted many researchers to look for ways to create a gap between them. Among the methods proposed are: confining the charge to the surface in systems composed of multi-layered graphene \cite{Morozov2005}, deformation of the graphene sheet \cite{Castro2009, Haugen2008, Ni2008, Huang2009, Jellal2020,Jakub}, application of different fields like electric, magnetic, or laser fields
\cite{Jellal2011, Mekkoui2021, biswas2012, biswas2013, Elaitouni2022,Morima}, and confinement of electron in graphene quantum dot \cite{hewage,Giavaras,Freitag}.

Laser technology has become a powerful tool for the advancement of investigations into graphene. Indeed, several studies have been conducted on the effect of a linear or circular laser field on the mobility of Dirac fermions in graphene \cite{Lasereffect,Sergy2011,laser,Lasereffect}, demonstrating that the laser can create a band gap in the energy spectrum. 
On the other hand, the tunneling effect has been studied for Dirac fermions in graphene subjected to a linear vector potential \cite{Jellal2011, MEKKAOUI2018}. Here, the infinite mass boundary condition reduced our 2D Dirac equation to a massively effective 1D Dirac equation with an effective mass equal to the quantized transverse momentum. 
Note that the transmission probabilities of Dirac fermions through different barriers are studied. In particular, rectangular barriers \cite{Jellal2011} and inclined barrier  \cite{MEKKAOUI2018}  all show that the transmission is almost equal to 1 for all values of incident energy, even if this energy is lower than the height of the barrier, which is called the Klein tunnel effect \cite{Klien,klien2}. 
 Barrier potentials in conjunction with laser fields have also been investigated \cite{biswas2016,biswas2012,biswas2013} and potential oscillations over time \cite{Collado2013,Jellal2019}. As a result, it was discovered that side bands $l \hbar \omega \ (l = 0, \pm 1, \cdots)$ were added to the energy spectrum, giving rise to an infinite number of transmission probabilities.

We investigate how Dirac fermions in graphene can pass through a slanted barrier potential in the presence of a laser field. The Dirac equation is solved using Floquet theory, and the energy spectrum is then obtained. The transfer matrix approach and boundary conditions enable us to ascertain the transmission probabilities corresponding to all energy bands. Under different conditions, only the central band $l=0$ and the two first side bands $l=\pm 1$ are examined numerically. It has been found that the amplitude and shape of the transmission probabilities are affected by the laser field and potential. In reality, a variety of oscillations are seen inside the barrier, the Klein effect is still evident, and the transmission of all bands is practically nonexistent.

This paper is organized as follows.  In Sec. \ref{TMM}, we set the mathematical formalism based on the Hamiltonian describing the present system. By solving the eigenvalue equation, we will be able to explicitly determine the eigenspinors in all regions composing the system under consideration. These solutions will allow us, in Sec. \ref{TTTT}, to use the boundary conditions at interfaces, and by employing the transfer matrix approach, we will calculate all transmission channels resulting from the oscillating barrier, which causes the energy subbands  $l \hbar \omega \ (l=0, \pm 1, \ldots)$. In Sec. \ref{NNNN}, we will numerically analyze our results by plotting the three first transmission probabilities as a function of the physical parameters. Finally, we conclude our work.

\section{Theoretical mode}\label{TMM}

\begin{figure}[H]
    \centering
    \includegraphics[scale=0.25]{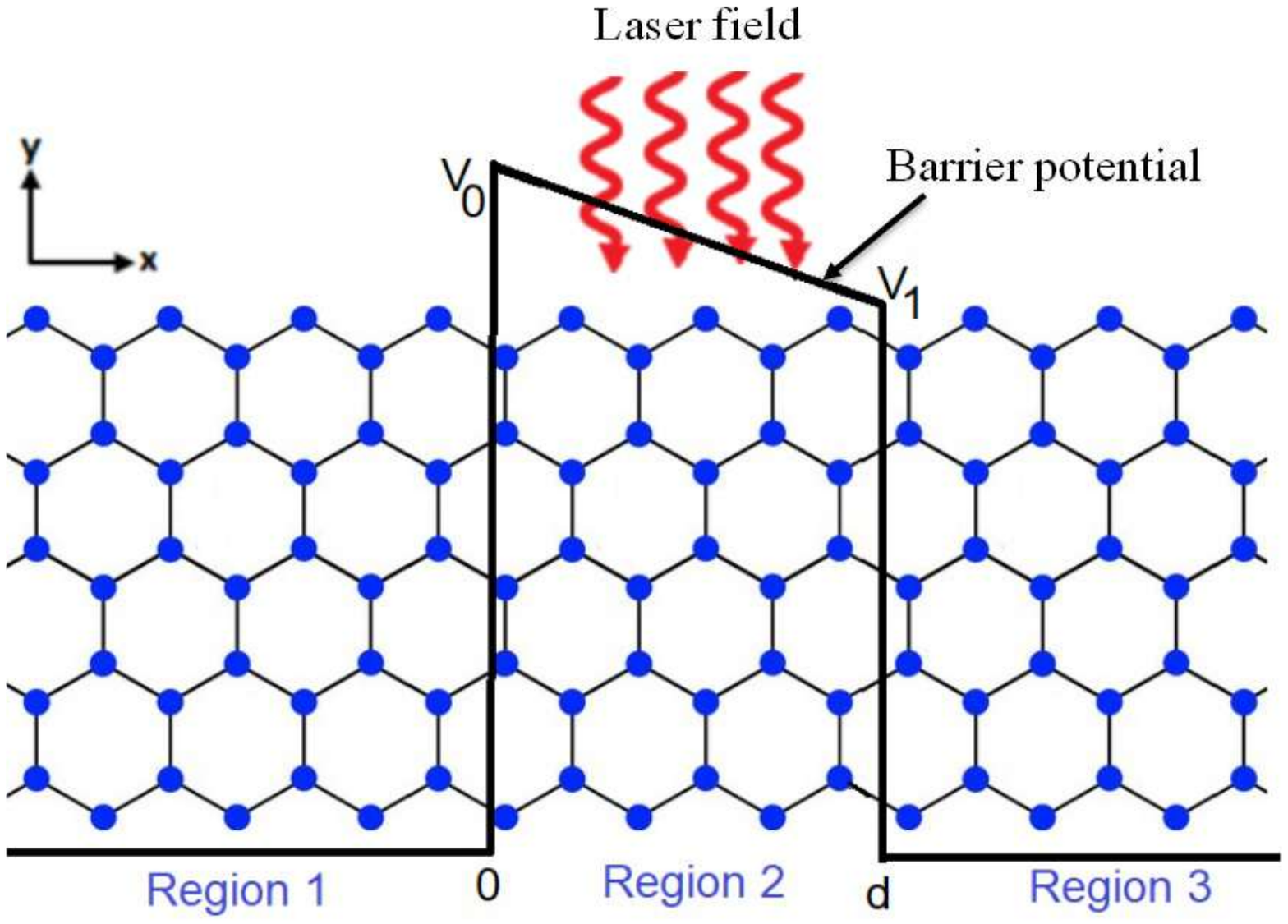}
    \caption{Schematic diagram of a graphene sheet electrostatic inclined barrier irradiated by a monochromatic laser field over a finite region of length $d$.}
    \label{fig1}
\end{figure}

We consider a graphene sheet subjected to a tilted potential $V(x)$ and irradiated by a linear polarization monochromatic laser $A(t)$ over a finite region of length $d$, while the other two regions are pure graphene as depicted in Fig. \ref{fig1}.

The Hamiltonian describing the system is given by
\begin{equation}\label{H1}
H=v_F\vec{\sigma}\cdot \left[\vec{p}-e\vec{A}(t)\right]+V(x)\mathbb{I}_2
\end{equation}
where $v_F$ is the Fermi velocity $\approx10^6 m/s$, $\sigma_{x,y}$ are the Pauli matrices, $\vec{p}=-i\hbar(\partial x, \partial y)$ is two dimensional momentum, $(-e)$ being the electron charge and $\mathbb{I}_2$ is $2\times 2$ identity matrix.
The scalar potential $V(x)$ varies along the $x$-axis and is represented by 
\begin{align}
    V(x)=\begin{aligned}
        & \begin{cases} -\beta x+ V_0 , & 0<x<d\\
            0, & \text{otherwise}
        \end{cases}
    \end{aligned}
\end{align}
which can be produced by two identical plates located at $x = 0 $ and $x = d$,
where $\beta=\frac{V_0- V_1}{d} $ and $V_0>V_1$. In the dipole approximation \cite{approx}, the vector potential $\vec A(t)$ of the laser field is given by 
 \begin{align}
   \vec{A}(\vec{r})=(0,A_0\cos(\omega t),0)
 \end{align}
where  $A_0=\frac{F}{ \omega}$ is the laser field amplitude. An electric field $\vec E$ with frequency $\omega$ and amplitude $F$ can generate such a potential. Indeed, we can write
\begin{equation}
    \vec{E}=-\frac{\partial }{\partial t}\vec{A}(t)=\vec{F} \sin(\omega t).
\end{equation}

The present system is composed of three regions labeled $j=1,2,3$. Then we have to determine the solutions of the energy spectrum in each region using the eigenvalue equation 
\begin{equation}\label{2HH}
    \left[  v_F\sigma_x p_x+ v_F\sigma_y(p_y-e A_y(t))+V_j(x)\mathbb{I}_2\right]\Psi_j(x,y,t)=i\hbar \frac{\partial}{\partial t}\Psi_j(x,y,t).
\end{equation}
Since the  system has a finite width $R$,  then one can use the  boundary conditions at $y = 0,R$
\begin{align}
&\Psi_1(x,0,t )=-\Psi_2(x,0,t)\\
&   \Psi_1(x,R,t)=-\Psi_2(x,R,t)
\end{align}
 to quantize the transverse wave vector 
\begin{align}
k_y=\frac{\pi}{R}\left(n+\frac{1}{2}\right), \quad  n=0,\pm1,\cdots.
\end{align}
The laser field is considered to oscillate periodically in time, which has an effect on the behavior of the wave function and the energy band. Consequently,  the Floquet approximation \cite{Floquet} should be taken into account to determine the solutions of the eigenvalue equation. The eigenspinors can also be separated in coordinates because the Hamiltonian is invariant along the $y $-direction. Combining all of these, we can write the eigenspinors solution of \eqref{2HH} as 
%
\begin{equation}\label{3HH}
\Psi_j(x,y,t)=(\psi_{j1}(x),\psi_{j2}(x))^Te^{ik_yy}\phi_0(t)e^{ -\frac{iEt}{\hbar}}
\end{equation}
where $E$ denotes the Floquet energy and $\phi_0(t)$ varies over time, $\phi_0(t+\tau)=\phi_0(t)$, $\tau$ denotes the laser field period and  $T$ denotes transpose. 
As a result,  \eqref{2HH} provides us with
\begin{align}
&\left(-i\frac{\partial}{\partial x}-i\left(k_y-\frac{F}{\omega}\cos(\omega t)\right)\right)\psi_{j2}(x)\phi_0(t)=i\psi_{j1}(x)\frac{\partial \phi_0(t)}{\partial t} +(E-V(x))\psi_{j1}(x)\phi_0(t)\label{4H}\\
&\left(-i\frac{\partial}{\partial x}+i\left(k_y-\frac{F}{\omega}\cos(\omega t)\right)\right)\psi_{j1}(x)\phi_0(t)=i\psi_{j2}(x)\frac{\partial \phi_0(t)}{\partial t} +(E-V(x))\psi_{j2}(x)\phi_0(t)\label{5H}
\end{align}
where  dimensionless quantities are introduced $A=\frac{A e d_0}{\hbar}$, $x=\frac{x}{d_0}$, $k_y=k_y d_0$, $V(x)=\frac{V(x)d_0}{\hbar v_F}$, $E=\frac{E d_0}{\hbar v_F}$, $t=\frac{t v_F}{d_0}$ and $\omega=\frac{\omega d_0}{v_F}$, with the length scale $d_0=\sqrt{\frac{\hbar v_F}{e F}}$.

It is worth noting that there are only two differential equations and three unknown wave functions, $\psi_{j1}(x)$, $\psi_{j2}(x)$, and $\phi_0(t)$. To overcome this situation, we apply the iterative method as a first approximation by supposing that $\psi_{21}$ and $\psi_{22}$ satisfy the Dirac equation in region 2 without laser irradiation. As a result, (\ref{4H}-\ref{5H}) becomes
\begin{align}
&\frac{F}{w}\cos(\omega t)\psi_{22}(x)\phi_0(t)=\psi_{21}(x)\frac{\partial }{\partial t}\phi_0(t)\label{4}\\
&-\frac{F}{w}\cos(\omega t)\psi_{21}(x)\phi_0(t)=\psi_{22}(x)\frac{\partial }{\partial t}\phi_0(t)\label{5}.
\end{align}
From these we end up with
the second order differential equation
\begin{equation}
\frac{\partial^2 \phi_0(t)}{\partial t^2}+\omega \tan(\omega t) \frac{\partial \phi_0(t)}{\partial t}+\frac{F^2}{\omega^2}\cos^2(\omega t) \phi_0(t)=0\label{6}
\end{equation}
showing the solution \cite{biswas2013}
\begin{align}
\phi_0(t)\approx e^{-i\frac{F}{\omega^2}\sin(\omega t)}.
\end{align}
Injecting this into \eqref{4H} to get
 \begin{align}
\left\{\left(-i\frac{\partial}{\partial x}-i\left(k_y-\frac{\partial}{\partial t}\right)\right)\psi_{22}(x)-\left(i\frac{\partial }{\partial t}+E-V(x)\right)\psi_{21}(x)\right\}e^{-i\frac{F}{\omega^2}\sin(\omega t)}e^{-im\omega t}=0\label{9H}.
 \end{align}
Generating the Bessel function $ J_m $ using the Jacobi-Anger expansion 
\begin{align}
    e^{-i\frac{F}{\omega^2} \sin(\omega t)}=\sum_{m=-\infty}^{+\infty} J_m\left(\frac{F}{\omega^2}\right)e^{-im\omega t}
\end{align}
and then we show that \eqref{9H} can be reduced to
\begin{align}
    \sum_{m=-\infty}^{+\infty} \left\{\left(-i\frac{\partial}{\partial x}-i\left(k_y-\frac{\partial}{\partial t}\right)\right)\psi_{22}(x)-\left(i\frac{\partial }{\partial t}+E-V(x)\right)\psi_{21}(x)\right\}J_m\left(\frac{F}{\omega^2}\right)e^{-im\omega t}=0\label{10}.
\end{align}
Consequently, we end up with
\begin{eqnarray}
\left(V(x)-(E+l\omega)\right)\psi_{21}(x)=-i\left(\frac{\partial}{\partial x} +(k_y-l\omega)\right)\psi_{22}(x)\label{15}.
\end{eqnarray}
Similarly, we get from
  \eqref{5H} 
\begin{eqnarray}
    \left(V(x)-(E+l\omega)\right)\psi_{22}(x)=-i\left(\frac{\partial}{\partial x}-(k_y-l\omega)\right)\psi_{21}(x)\label{16}
\end{eqnarray}
Let us make a change to the spinor components to obtain the Schrödinger equation. This is 
\begin{align}
&\psi_{21}(x)=\chi_1+i \chi_2\\
&	\psi_{22}(x)=\chi_1-i \chi_2
\end{align}
After substitution in (\ref{15}-\ref{16}), we obtain
\begin{equation}\label{13A}
\frac{\partial^2\chi_1}{\partial x^2}+\left(-i\frac{\partial V(x)}{\partial x}+(k_y-l\omega)^2+\left(V(x)-(E+l\omega)\right)^2\right)\chi_1=0
\end{equation}
which can be written as Weber's differential equation \cite{math}
\begin{equation}\label{17A}
\frac{\partial^2\chi_1(z_l)}{\partial {z_l}^2}+\left(\frac{1}{2}-v_l-\frac{z^2_l}{4}\right)\chi_1(z_l)=0
\end{equation}
where we have set
\begin{align}
&z_l=\sqrt{\frac{2}{\beta}}e^{\frac{i\pi}{4}}(\beta x+E_l)\\
&   v_l=\frac{i}{2\beta}\left(k_y-l\omega\right)^2\\
& E_l=E+l\omega-V_0.	
\end{align}
The solution of \eqref{17A} is given by
\begin{eqnarray}
\chi_1(z_l) = C_{1,l}D_{v_l-1}(z_l)+C_{2,l}D_{-v_l}(-z^*_l)
\end{eqnarray}
where $ C_{1,l} $ and  $ C_{2,l}$ are two constants.
From $\chi_1$, we derive the second component
 \begin{eqnarray}
\chi_2(z_l) =C_{1,l}\frac{\sqrt{2\beta}e^{-i\frac{\pi}{4}}}{k_y-l\omega}D_{v_l}(z_l)+\frac{C_{2,l}}{(k_y-l\omega)}\left(2\left(\beta x +E_l\right)D_{-v_l}(-z^*_l)+\frac{\sqrt{2B}e^{i\frac{\pi}{4}}}{k_y-l\omega}D_{1-v_l}(-z^*_l)\right).
\end{eqnarray}
These allow us  to express  $\psi_{21}(x)$ and $\psi_{22}(x)$
as follows
\begin{eqnarray}
\psi_{21}(x)=C_{1,l}\mu^+_l(x)+C_{2,l}\zeta^+_l(x)\\
\psi_{22}(x)=C_{1,l}\mu^-(x)+C_{2,l}\zeta^-(x)
\end{eqnarray}
where we shave defined
\begin{eqnarray}
\mu_l^{\pm}(x)&=&D_{v_l-1}(z_l)\pm \frac{\sqrt{2\beta}}{k_y-l\omega}e^{\frac{i\pi}{4}}D_{v_l}(z_l)\\
\zeta_l^{\pm}(x)&=&\frac{1}{k_y-l\omega}\left(\left(k_y-l\omega\mp(2i\beta x+2iE_l)\right)D_{-v_l}(-z^*_l)\pm\sqrt{2\beta}e^{-i\frac{\pi}{4}}D_{1-v_l}(-z^*_l)\right).
\end{eqnarray}
Combining all to write the eigenspinors 
 in  region 2  as
\begin{equation}
\Psi_{2}(x,y,t)=e^{ik_yy}\sum_{l=-\infty}^{+\infty}\left[C_{1,l}\begin{pmatrix}
\mu^+_l(x)\\\ \mu^-_l(x)
\end{pmatrix}
+C_{l,2}\begin{pmatrix}
\zeta^+_l(x)\\ \zeta^-_l(x)
\end{pmatrix}\right]\sum_{m=-\infty}^{+\infty}J_{m}\left(\frac{F}{\omega^2}\right)e^{-i(E+l\omega+m\omega)t}.
\end{equation}

In the regions 1 and 3 have only pristine graphene, the spinors in both regions written as follows \cite{MEKKAOUI2018}
\begin{align}
&\Psi_1(x,y,t)=e^{ik_yy}\left[\delta_{l,0}\begin{pmatrix}
1\\ \Lambda_l
\end{pmatrix}e^{ik_lx}e^{-iEt}+\sum_{l=-\infty}^{+\infty}r_l\begin{pmatrix}
1\\-\Lambda^*_l
\end{pmatrix}e^{-ik_lx}e^{-i(E+l\omega)t}\right]\sum_{m=-\infty}^{+\infty}J_m\left(\frac{F}{\omega^2}\right)e^{-im\omega t}
\\
&
\Psi_{3}(x,y,t)=e^{ik_yy}\sum_{m=-\infty}^{+\infty}\left[t_l\begin{pmatrix}
1\\ \Lambda_l
\end{pmatrix}e^{ik_lx}+b_l\begin{pmatrix}
1\\-\Lambda^*_l
\end{pmatrix}e^{-ik_lx}\right]\delta_{m,l}e^{-i(E+m\omega)t}
\end{align}
corresponding to the eigenvalues 
\begin{align}
	E+l\omega=s_l\sqrt{k^2_l+k^2_y}
\end{align}
with  the reflection $r_l$ and transmission $t_l$ amplitudes,
$
	\Lambda_l 
	=s_le^{i\phi_l}$,
$
	\phi_l=\tan^{-1}\frac{k_y}{k_l}
$,
 $\delta_{m,l}=J_{m-l}(0)$,  $s_l=\text{sgn}(E+l\omega)$
and ${b_l}$  is null vector.

\section{Transmission probabilities}\label{TTTT}

To determine the transmission and reflection amplitudes, we use 
the boundary conditions  at the interfaces $x=0$ and $ x=d$ 
\begin{align}
&	\Psi_1(0,y,t)=\Psi_{2}(0,y,t)\\
&	\Psi_{2}(d,y,t)=\Psi_{3}(d,y,t)
\end{align}
together with the orthogonality of $e^{im\omega t}$ to end up with
 four unknown parameters, each one there in an infinity of modes
\begin{align}
&\delta_{m,0}+r_m=\sum_{l=-\infty}^{+\infty}\left(c_{1,l}\mu^+_l(0)+c_{2,l}\zeta^+_l(0)\right)J_{m-l}\left(\frac{F}{\omega^2}\right)\\
&\delta_{m,0}\Lambda_m-r_m\Lambda_m^*=\sum_{l=-\infty}^{+\infty}\left(c_{1,l}\mu^-_l(x)(0)+c_{2,l}\zeta^-_l(0)\right)J_{m-l}\left(\frac{F}{\omega^2}\right)\\
&t_me^{ik_md}+b_me^{-ik_md}=\sum_{l=-\infty}^{+\infty}\left(c_{1,l}\mu^+_l(d)+c_{2,l}\zeta^+_l(d)\right)J_{m-l}\left(\frac{F}{\omega^2}\right)\\
&t_m\Lambda_me^{ik_md}-b_m\Lambda_m^*e^{-ik_md}=\sum_{l=-\infty}^{+\infty}\left(c_{1,l}\mu^-_l(d)+c_{2,l}\zeta^-_l(d)\right)J_{m-l}\left(\frac{F}{\omega^2}\right).
\end{align}
These relations can be expressed using the transfer matrix approach to connect the three regions of the system under consideration. This is 
\begin{equation}\label{35}
\begin{pmatrix}
\Gamma_1\\
\Gamma'_1
\end{pmatrix}
=\mathbb{M} \begin{pmatrix}
\Gamma_2\\
\Gamma'_2
\end{pmatrix}
\end{equation}
with the total transfer matrix
\begin{align}
\mathbb{M}=\begin{pmatrix}
	M_{1,1}&    M_{1,2}\\
	M_{2,1}&    M_{2,2}
\end{pmatrix}=\mathbb{M}(0,1)\cdot \mathbb{M}(1,2)	
\end{align}
where $\mathbb{M}(j,j+1)$ is  the transfer matrix connecting  regions $j$ and $j+1$
\begin{align}
&\mathbb{M}(0,1)=\begin{pmatrix}
\mathbb{I}_2 & \mathbb{I}_2 \\
\mathbb{S}^+&\mathbb{S}^-\\
\end{pmatrix}^{-1}
\begin{pmatrix}
\mathbb{F}^+_0&\mathbb{R}^+_0\\
\mathbb{F}^-_0&\mathbb{R}^-_0
\end{pmatrix}
\\
&
\mathbb{M}(1,2)=\begin{pmatrix}
\mathbb{F}^+_d&\mathbb{R}^+_d\\
\mathbb{F}^-_d&\mathbb{R}^-_d
\end{pmatrix}^{-1}
\begin{pmatrix}
\mathbb{I}_2 & \mathbb{I}_2 \\
\mathbb{S}^+&\mathbb{S}^-\\
\end{pmatrix}
\begin{pmatrix}
\mathbb{Q}^+&\mathbb{O} \\
\mathbb{O}&\mathbb{Q}^-\\
\end{pmatrix}
\end{align}
and the quantities
\begin{align}
\mathbb{S}^{\pm}=\pm\delta_{m,l}\Lambda_l^{\pm1},  \quad 
\mathbb{F}^\pm_z=\mu_l^\pm(z)J_{m-l}\left(\frac{F}{\omega^2}\right)\\
\mathbb{R}^\pm_z=\zeta_l^\pm(z)J_{m-l}\left(\frac{F}{\omega^2}\right),  \quad
\mathbb{Q}^\pm=e^{\pm ik_lL}\delta_{m,l}
\end{align}
where $\mathbb{O}$ is  the zero matrix, $\mathbb{I}_2$ is the unit matrix and $z=\{0,d\}$.
We consider the fermions moving from left to right with the energy $E$, then from Eq. \eqref{35} we get
\begin{equation}
\Gamma_2=\mathbb{M}^{-1}_{1,1}\Gamma_1
\end{equation}
where $\Gamma_1=\delta_{0,l}$ is the Kronecker symbol and $\Gamma_2=t_l$.
Because $n,l$ varies from $-\infty $ to $+\infty $, the transfer matrix approach, which we determined above, is of infinite order, making it difficult to solve. To accomplish this, we replace the infinite series with a finite number of terms ranging from $-N$ to $N$, where $N>\frac{F}{\omega^2}$, and we limit ourselves only to the first energy side bands $E\pm m\omega$, yielding the result 
\begin{equation}
t_{-N+k}=M'[k+1,N+1]
\end{equation}
with $M'=M^{-1}_{1,1}$ and $k=0, 1, 2, \cdots N$.

On the other hand, the transmission probability can be expressed by using the continuity equation. According to our Hamiltonian, the general expression of the current density takes the following form:
\begin{align}
J=\Psi_j^{\bb\dagger}(x,y,t)\sigma_x\Psi_j(x,y,t)	
\end{align} 
 which makes it possible to determine the current densities in each region $j$. There are given by
\begin{align}
&J_{in,0}=(\Lambda_0+\Lambda^*_0)\label{47}\\
&J_{re,l}=r^*_lr_l(\Lambda_l+\Lambda^*_l)\label{48}\\
&J_{tr,l}=t^*_lt_l(\Lambda_l+\Lambda^*_l)\label{49}.
\end{align}
Then by using the relations
\begin{equation}
T_l=\frac{|J_{tr,l}|}{|J_{in,0}|},\quad R_l=\frac{|J_{re,l}|}{|J_{in,0}|}
\end{equation}
one can  express of the transmission $T_l$ and reflection $R_l$ probabilities as
\begin{eqnarray}
T_l=s'\frac{\cos{\theta_l}}{\cos{\theta_0}}|t_l|^2,\quad R_l=s'\frac{\cos{\theta_l}}{\cos{\theta_0}}|r_l|^2
\end{eqnarray}
where  the two angles are
\begin{align}
\cos{\theta_l}=\frac{k_l}{\sqrt{k_l^2+k_y^2}}, \quad
 	\cos{\theta_0}=\frac{k_0}{\sqrt{k_0^2+k_y^2}}
\end{align}
and the sign is 
$s'=s_ls_0$.

Because of the difficulty of numerical calculation, we limit our study to the three first bands: the central band  $l=0$ (zero photon exchange) and the two first side-bands $l=\pm1$ (absorption and emission)
\begin{align}
& t_{-1}=M'[1,2]\\
& t_{0}=M'[2,2]\\
& t_{1}=M'[3,2].
\end{align}
In the following section, we will plot the transmission probabilities corresponding to $l =0, \pm1$ as a function of  physical parameters characterizing our system.

\section{Numerical results} \label{NNNN}

\begin{figure}[H]
	\centering
	\subfloat[]{
		\centering
		\includegraphics[scale=0.5]{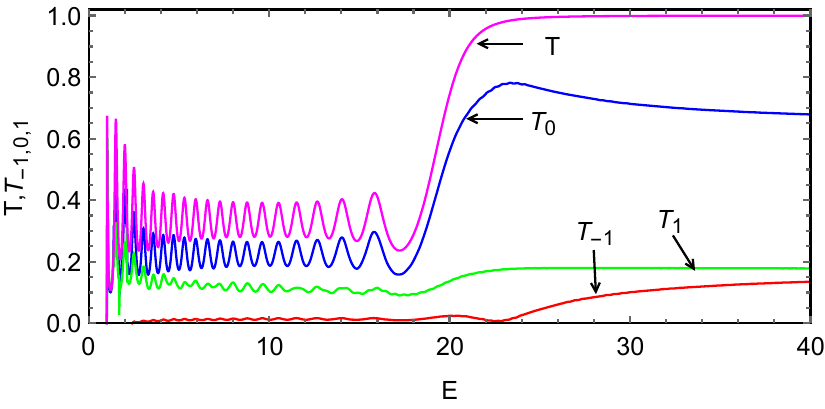}
		\label{fig2a}
	}\subfloat[]{
		\centering
		\includegraphics[scale=0.5]{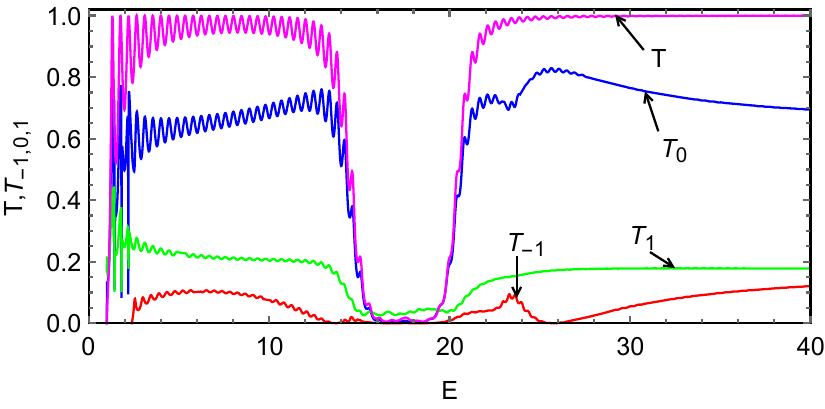}
		\label{fig2b}}
	\caption{{(Color online) Transmission probabilities as a
			function of the incident energy $E$ with $d=7$, $V_{0}=20$,
			$k_{y}=1$, $\omega=1.4$, $F=0$ for $T$ (magenta line), $F=0.9$ for the
			central transmission band $T_{0}$ (blue line) and two first
			sidebands $T_{-1}$ (red line), $T_{1}$ (green line). (a):
			$V_{1}=0$ and (b): $V_{1}=15$.}}
	\label{fig2}
\end{figure}

 Fig. \ref{fig2} represents the transmission probabilities as a function of the incident energy $E$. The total transmission $T=T_0$ (magenta line) is shown in Fig. \ref{fig2a}, and the other bands $T_{\pm 1}$ are zero in the absence of laser irradiation $F=0$ of a linear barrier ($V_0 = 20$, $V_1 = 0$). Let us now examine what happens in the intermediate region if laser irradiation $F=0.9$ is introduced. This is shown by the transmission for the central band $T_0$ (blue line) and the two first sidebands $T_{-1}$ (red line) and $T_1$ (green line). We  find that the laser irradiation is very significant in determining the relative transmission probabilities of two first sidebands. Transmissions are prohibited for the energy $E=\pm\omega$, resulting in a change in $k_y$ to $k_y\pm\omega$, which serves as an effective mass $m^*$ \cite{Jellalmass2012, Jellal2019}. The total transmission and the transmission of the central band oscillate around a minimum for the energies $E<V_0$, then increase passing by $E=V_0$ up to a maximum value, and the total transmission (the sum of the three) tends towards unity (Klein tunneling), indicating that our system exhibits classic behavior after the barrier, indicating the presence of the Klein effect. 
 The same transmissions are plotted in Fig. \ref{fig2b}, but with a two-potential barrier $(V_0 = 20, V_1 = 15)$. The transmission of the central band is more dominant in the energy zone $k_y<E<V_1$, and the total transmission oscillates in the neighborhood of unity, showing that the Klein paradox is always present. We observe a total reflection in the energy zone $V_1<E<V_0$, with a very small transmission for the band corresponding to $l=1$. In the third zone, $E>V_1$, total transmission rapidly increases towards unity, and $T_0$ increases with an apparent oscillation. Finally, we see that changing the barrier's inclinison coefficient, $\beta=\frac{V_0-V_1}{d}$, affects the shape of the transmissions and the number of oscillations. 
 
 

\begin{figure}[H]
	\centering
	\subfloat[]{
		\centering
		\includegraphics[scale=0.5]{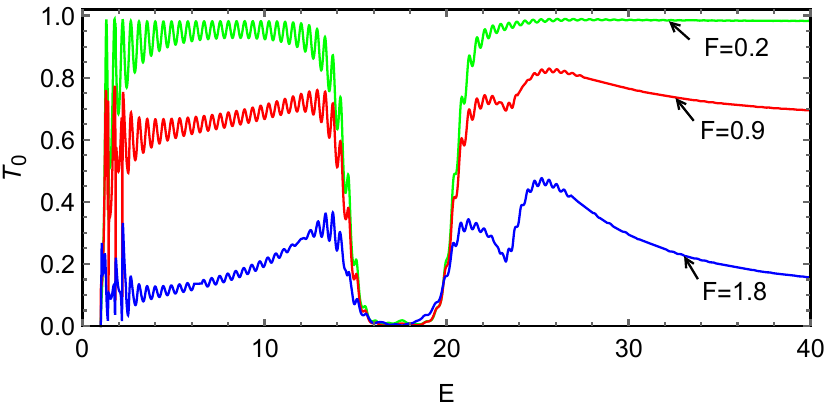}
		\label{fig3a}
	}\subfloat[]{
		\centering
		\includegraphics[scale=0.5]{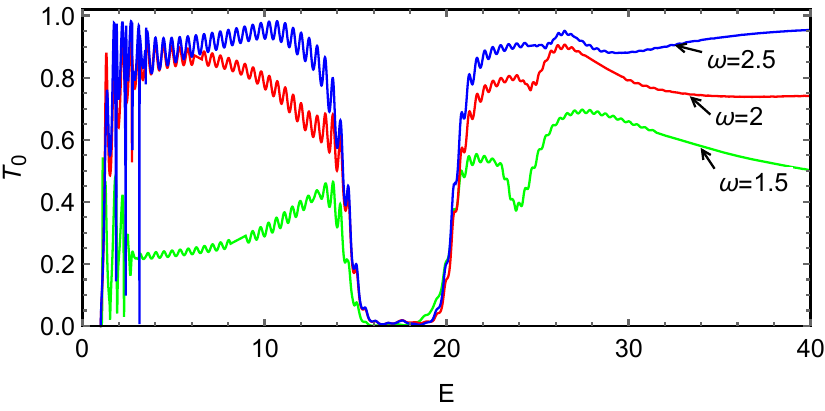}
		\label{fig3b}}
	\caption{{(Color online) Transmission probability  for the
			central band $T_{0}$ as a function of the incident energy $E$ with
			$d=7$, $V_{0}=20$, $V_{1}=15$, $k_{y}=1$. (a): $\omega=1.4$, $F=0.2$
			(green color), $F=0.9$ (red line) and $F=1.8$ (blue line). (b):
			$F=1.8$, $\omega=1.5$ (green color), $\omega=2$ (red line) and $\omega=2.5$ (blue
			line).}}
	\label{fig3}
\end{figure}

Fig. \ref{fig3} illustrates  the transmission of the central band ($l=0$) as a function of the incident energy $E$ with $d=7$, $V_0=20$, $V_1=15$, $k_y=1$ for $\omega=1.4$, $F=\{0.2, 0.9, 1.8\}$ in Fig. \ref{fig3a} and $F=1.8$, $\omega=\{1.5, 2, 2.5\}$ in Fig. \ref{fig3b}. 
We see that the transmission probability with zero photon exchange varies in an oscillatory way, and it has the same allure. However, with an attenuation proportional to laser amplitude $F$ and frequency $\omega$, oscillations appear after the barrier. When the amplitude of the laser field $F$ is increased, the oscillation becomes important, and the amplitude of $T_0$ decreases because the magnetic field suppresses the transmission without photon exchange and increases the transmission with exchange, as shown in Fig. \ref{fig3a}. But because the variation of $\omega$  has an opposite effect than $F$, the  transmission with zero photon exchange becomes important if the frequency decreases with the increase in oscillations, as depicted in Fig. \ref{fig3b}. Finally, the ratio $\frac{F}{\omega^2}$ has an important role because it determines the shape, the number of oscillations, and the amplitude of the transmission. We always have an anti-Klein tunnel inside the barrier for any values of $\omega$ and $F$.

\begin{figure}[H]
        \centering
        \subfloat[]{
            \centering
            \includegraphics[scale=0.5]{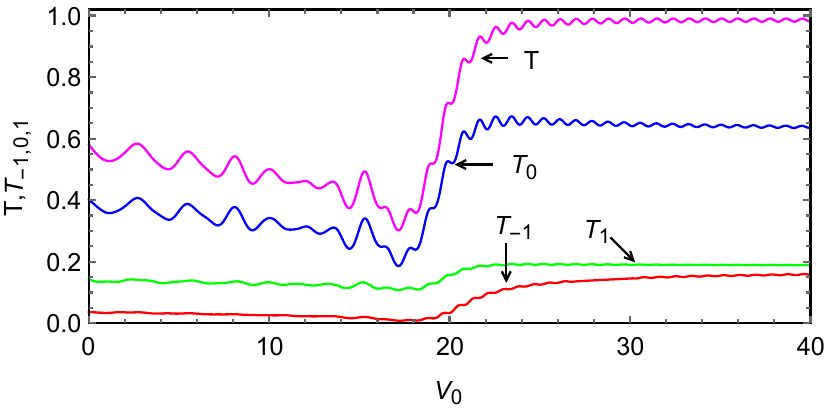}
            \label{fig4a}
        }\subfloat[]{
            \centering
            \includegraphics[scale=0.5]{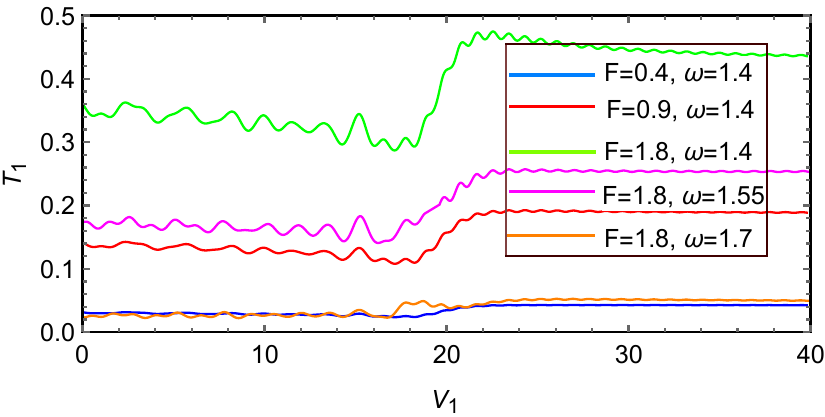}
            \label{fig4b}}
\caption{{(Color online) Transmission probabilities as a
function of potential $V_1$ with $d=7$, $V_{0}=40$, $E=20$,
$k_{y}=1$. (a): $w=1.4$, $F=0.9$ for the total
transmission $T$ (magenta line), the central transmission band
$T_{0}$ (blue line) and two first sidebands $T_{-1}$ (red line),
$T_{1}$ (green line). (b): Transmission probability  $T_{1}$ with
$\{F=0.4, \omega=1.4\}$ (blue line), $\{F=0.9, \omega=1.4\}$ (red
line), $\{F=1.8, \omega=1.4\}$ (green line), $\{F=1.8,
\omega=1.55\}$ (magenta line), $\{F=1.8, \omega=1.7\}$ (orange
line).}}
    \label{fig4}
\end{figure}

Fig. \ref{fig4} depicts transmissions as a function of the barrier  potential $V_1$ under various conditions. Taking $ \omega = 1.4 $ and $ F = 0.9 $ in Fig. \ref{fig4a}, 
 we can see that transmissions decrease from zero to $E - 2k_y$, then increase to $E + 2k_y$. When $V_1>E+2k_y$, the transmissions are maximized, and the total transmission is almost equal to unity, i.e., the Klein tunneling. Fig. \ref{fig4b} represents the transmission $T_1$ as a function of $V_1$ with different values of $\omega$ and $F$. We observe that when $F$ increases without changing the frequency, $T_1$ (transmission with photon exchange) increases.
	Now, if we fix $F$ and vary the frequency, we observe the opposite behavior. In other worlds, we can characterize the field by the ratio {$\frac{F}{\omega^2}$}. As a result, if this ratio increases, $T_1$  increases too, and vice verse.

\begin{figure}[H]
	\centering
	\subfloat[]{
		\centering
		\includegraphics[scale=0.5]{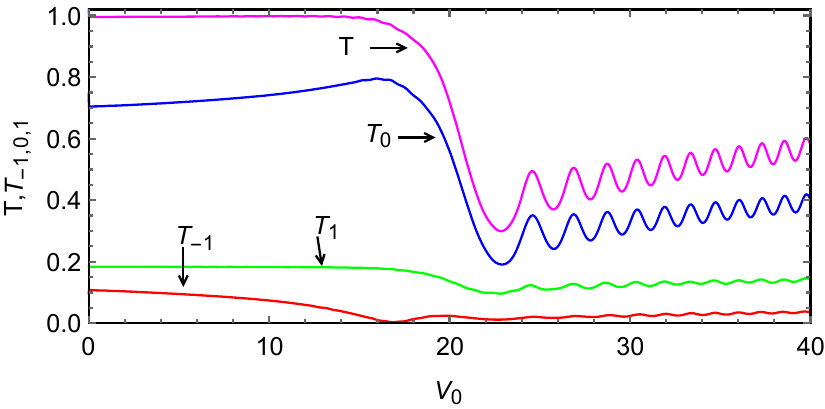}
		\label{fig5a}
	}\subfloat[]{
		\centering
		\includegraphics[scale=0.5]{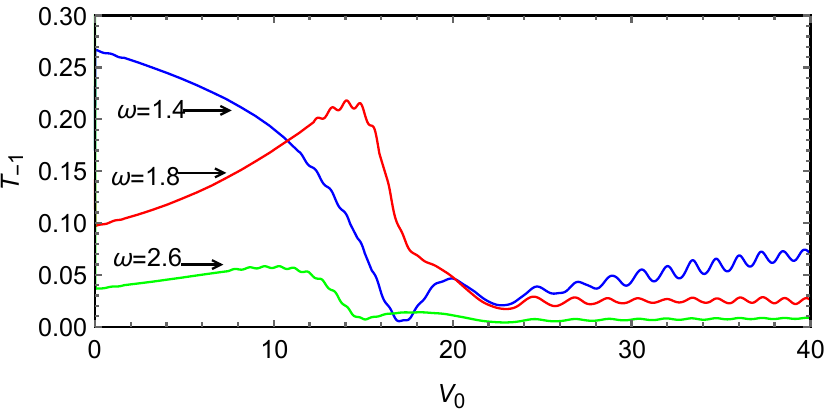}
		\label{fig5b}}
	\caption{{(Color online) Transmission probabilities as a
			function of potential $V_0$ with $d=7$, $V_{1}=0$, $E=20$,
			$k_{y}=1$. (a): $\omega=1.4$, $F=0.9$ for total
			transmission $T$ (magenta line), the central transmission band
			$T_{0}$ (blue line) and two first sidebands $T_{-1}$ (red line),
			$T_{1}$ (green line). (b): Transmission probability  $T_{-1}$ with
			$F=1.8$, $\omega=1.4$ (blue line), $\omega=1.8$ (red line) and
			$\omega=2.6$ (green line).}}
	\label{fig5}
\end{figure}

Fig. \ref{fig5} shows the transmission probabilities as a function of the barrier potential $V_0$ with $d=7$, $V_1=0$, $E=20$ and $k_y=1$. When we use $\omega=1.4$ and {$\frac{F}{\omega^2}=0.46$} in Fig. \ref{fig5a},
we see that the transmission $T_0$ is always  dominant compared to $T_\pm1$. The total transmission $T$ is  almost equal to unity for $V_1<E-2k_y$ that is, for the Klein zone. In the interval $E-2k_y<V_0<E+2k_y$, it is clear that $T_0$ and $T$ decrease rapidly. After $E+2k_y$, the transmissions increase in phase by exhibiting an oscillatory behavior. 
In Fig. \ref{fig5b}, we plot the transmission probability with photon emission ($T_{-1}$) as a function of $V_1$ with different values of $\omega$. The amplitude and shape of {$T_{-1}$} clearly depend on  $F$ and $\omega$. Indeed, the amplitude is maximum and greater than $0.25$ for {$\frac{F}{\omega^2}=0.91$} (blue line), then it rapidly decreases to zero after having increased in an oscillatory manner. 
The amplitude increases to a maximum greater than $0.2$ for {$\frac{F}{\omega^2}=0.55$} (red line), then rapidly decreases to zero. For {$\frac{F}{\omega^2}=0.26$} (green line), the amplitude does not exceed $0.05$ then it decreases to zero. It can be concluded that the transmission probability  with photon emission increases if the laser field intensity increases, which suppresses the Klein tunneling.

\begin{figure}[H]
	\centering
	\subfloat[]{
		\centering
		\includegraphics[scale=0.5]{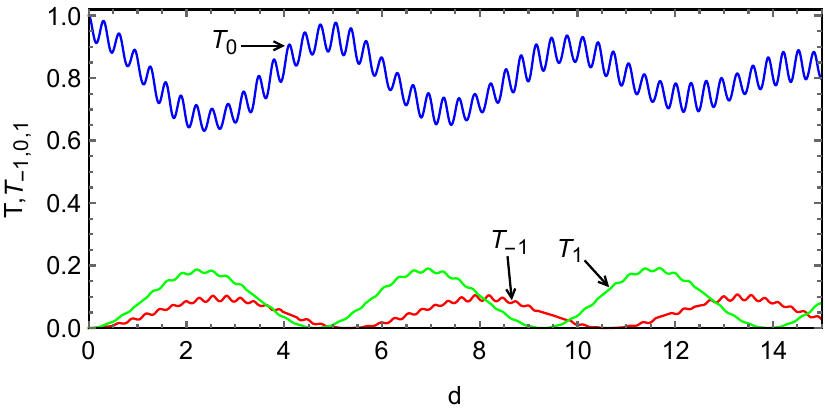}
		\label{fig6a}
	}\subfloat[]{
		\centering
		\includegraphics[scale=0.5]{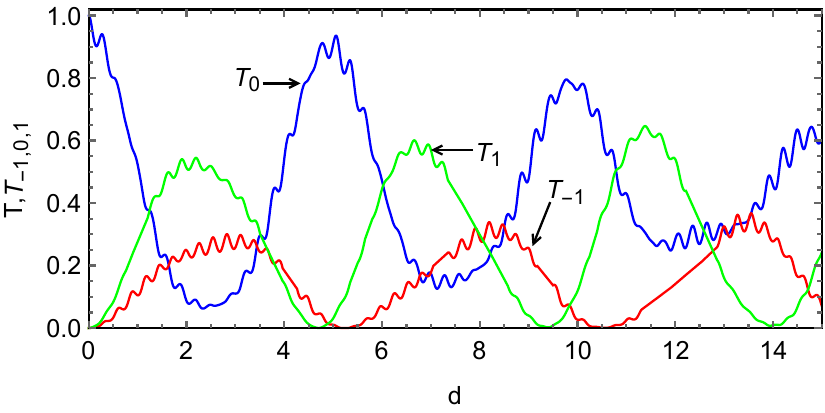}
		\label{fig6b}}
	\caption{{(Color online) Transmission probabilities for the
			central band $T_{0}$ (blue line) and two first sidebands $T_{-1}$
			(red line), $T_{1}$ (green line) as a function of the barrier
			width $d$ with $V_0=20$, $V_{1}=10$, $E=5$, $k_{y}=1$. $\omega=1.4$.
			(a): $F=0.8$ and (b): $F=1.8$.}}
	\label{fig6}
\end{figure}

The transmission probabilities for the first three bands are shown in Fig. \ref{fig6} as a function of barrier width $d$, with $V_0=20$, $V_1=10$, $E=5$, $k_y=1$, and $\omega=1.4$. 
 In Fig. \ref{fig6a}, for $F=0.9$, we observe that all the transmissions oscillate in a sinusoidal way with the same period but different amplitudes, with a large difference between $T_0$ and {$T_{\pm1} $}. In the presence of a laser field $F$, the flux of incidence fermions redistributes on all the sub-bands, but the process of transmission with zero photon exchange ($T_0$) is always dominant over the process of transmission with photon exchange ($T_{\pm1 }$). The transmission probability of the central band (with zero photon exchange) oscillates around unity. With the increase in the laser field, $ F = 1.8 $, the spacing between the transmissions disappears, and the transmission probability with photon exchange increases and can offset the transmission with zero photon exchange. The same thing happens when the barrier widths are increased, as shown in Fig. \ref{fig6b}. At very precise values of $d$, the transmissions with or without photon exchange are equal. As a conclusion, the increase in the laser field  or the barrier width motivates (increases) the process of transmission with photon exchange between the barrier and the incident region.

\bibliographystyle{plan}
\bibliography{ref.bib}
\section{Conclusion}

We theoretically investigated a system consisting of three graphene regions: $1,2,3$, virgin graphene, and an intermediate region with a laser field generated by an electric field of amplitude $F$ and frequency {\color{red}$\omega$} in the presence of an inclined scalar potential generated between two identical plates by two different potential generators $V_0, V_1$. First, we solved the eigenvalue equation to determine the wave functions corresponding to each region. For this, we used the Floquet theory and the solution of Weber differential equation. To determine the transmission probabilities, we employed the boundary conditions and used transfer matrix approach, which give an infinite order transfer matrix. To simplify, we limited our studies to the first three bands, that are the central band $l=0$ and the two first side bands $l=\pm1$.

We plotted the transmission probabilities as a function of the physical parameters $E$, $V_0$, $V_1$, $d$, and $\omega$. In summary, our numerical results showed that the laser field plays an important role in the barrier.  The oscillations of this barrier over time generate several modes of transmission probabilities. It is demonstrated that the reflection is complete inside the barrier for all modes. It is found that  the variation of the frequency $\omega$ or the amplitude $F$ (i.e. $\frac{ F}{\omega^2}$) changes the form and the amplitude of transmissions. More importantly, we observed that the transmission of the central band is always more dominant than the sidebands, and the Klein zone is always present, especially for energies lower than the barrier potential.

\end{document}